# Electrons surfing on a sound wave as a platform for quantum optics with flying electrons

Sylvain Hermelin[1], Shintaro Takada[2], Michihisa Yamamoto[2,3], Seigo Tarucha[2,4],

Andreas D. Wieck[5], Laurent Saminadayar[1,6], Christopher Bäuerle[1] and Tristan Meunier[1]

[1]*Institut Néel, CNRS and Université Joseph Fourier, 38042 Grenoble,*

*France* [2]*Department of Applied Physics, The University of Tokyo, Tokyo, 113-8656,*

*Japan,*

[3]*ERATO-JST, Kawaguchi-shi, Saitama 331-0012, Japan,*

[4]*ICORP (International Cooperative Research Project) Quantum Spin Information*

*Project, Atsugi-shi, Kanagawa, 243-0198, Japan,*

[5]*Lehrstuhl für Angewandte Festkörperphysik, Ruhr-Universität Bochum,*

*Universitätsstraße 150, 44780 Bochum, Germany*

[6]*Institut Universitaire de France, 103 boulevard Saint-Michel, 75005 Paris, France*

**Electrons in a metal are indistinguishable particles that strongly interact with other electrons and their environment. Isolating and detecting a single flying electron after propagation to perform quantum optics like experiments at the single electron level is therefore a challenging task. Up to date, only few experiments have been performed in a high mobility two-dimensional electron gas where the electron propagates almost ballistically [3, 4, 5]. Flying electrons were detected via the current generated by an ensemble of electrons and electron correlations were encrypted in the current noise. Here we demonstrate the experimental realisation of high efficiency single electron source and single electron detector for a quantum medium where a single electron is propagating isolated from the other electrons through a one-dimensional (1D) channel. The moving potential is excited by a surface acoustic wave (SAW), which carries the single electron along the 1D-channel at a speed of**

**3μm/ns. When such a quantum channel is placed between two quantum dots, a single electron can be transported from one quantum dot to the other, which is several micrometres apart, with a quantum efficiency of emission and detection of 96 % and 92 %, respectively. Furthermore, the transfer of the electron can be triggered on a timescale shorter than the coherence time $T_2$\* of GaAs spin qubits[6]. Our work opens new avenues to study the teleportation of a single electron spin and the distant interaction between spatially separated qubits in a condensed matter system.**

Quantum electron-optics is a field aiming at the realisation of photon experiments with flying electrons in nanostructures at the single electron level. Important tools to infer complex photon correlations inaccessible from ensemble measurements are single photon sources and single photon detectors. In opposition with photons, electrons are strongly interacting particles and they usually propagate in a Fermi sea filled with other electrons. Therefore each electron inevitably mixes with the others of the Fermi sea which implies that the quantum information stored within the charge or the spin of the single electron will be lost over short lengths. To perform quantum electron-optics experiments at the single electron level, one therefore needs a single electron source, a controlled propagating medium and a single electron detector. For charge degree of freedom, it has been proposed that edge states in the quantum Hall effect can serve as a 1D propagating channel. Due to Coulomb blockade, quantum dots have been demonstrated to be a good single electron source [7, 8] and can also serve as a single electron detector. Indeed, once an electron is stored in a quantum dot, its presence can be inferred routinely using charge detection [9]. Nevertheless, retrapping the electron in another quantum dot after propagation in an edge state turns out to be extremely difficult and presently all the information extracted from such experiments are coming from ensemble measurements [10, 11]. Here, we demonstrate that a single flying electron – an electron surfing on a sound wave - can be sent on demand from a quantum dot via a 1D quantum channel and retrapped in a second quantum dot after propagation. The 1D quantum channel consists of a several micrometre long depleted region in a two-dimensional electron gas (2DEG). The electron is dragged along by exciting a surface acoustic wave (SAW) and propagates fully isolated from the other electrons inside the 1D-channel[12]. Loading and unloading of the flying electron from the quantum channel into a quantum dot turn out to be highly efficient processes. Moreover, we demonstrate that the transfer of the electron can be triggered with a timescale smaller than the coherence time $T_2$* of GaAs spin qubits[6]. Since both electron spin directions are treated on the same foot in the SAW quantum channel, one expects that the spin coherence during the transport is conserved. Naturally, new possibilities will emerge to address the question of scalability in spin qubit systems[6, 14, 15].

To transport a single electron from one quantum dot to the other separated by a 3µm 1D-channel (see Fig. 1 and methods section), the following procedure is applied. First, the region between the two electrodes, which define the 1D channel is fully depleted. As a consequence, direct linear electron transport from one end of the channel to the other is blocked since the Fermi energy lies below the potential induced by the gates. Second, by applying a microwave excitation to the interdigitated transducer (IDT), SAW induced moving quantum dots are generated[12] due to the piezo-electric properties of GaAs (see also supplementary materials). Adding a quantum dot to each side of the 1D channel and tuning both quantum dots into the single electron regime, it is then possible to transport a single electron from one quantum dot across the 1D-channel and catch it inside the second quantum dot. Stability diagrams for both quantum dots as a function of the applied voltage on the two gates controlling the two barriers of the quantum dot are presented in Fig. 2a and Fig. 2b. They demonstrate that the system can be tuned into the few-electrons regime[16]. As expected, the charge degeneracy lines disappear when the barrier height between each dot and the reservoir is increased (corresponding to voltages on the gate Vb and Vb' more and more negative).

This also changes the position of the quantum dot minimum and brings the electron closer to the 1D-channel, to a position where a better transfer to SAW quantum dots is expected.

The protocol of the single electron source for a SAW quantum channel is a sequence made of three dot gate voltage steps (see Fig. 2a). At the working point A on Fig. 2a, the left quantum dot (the single electron source) is loaded with one electron on a timescale close to microseconds and unresolved with the set-up detection bandwidth. It is then brought rapidly to the working point B where the chemical potential of the single electron state lies above the Fermi energy and the coupling to the 1D-channel is expected to be large. The actual position of B is not very crucial as long as the electron is sufficiently protected from tunnelling out of the dot and the dot potential high enough to facilitate the charging of the electron into the moving SAW dot (see inset of Fig. 2a). For each sequence, the QPC conductance time-trace is recorded to observe single shot loading and unloading of the dot. This sequence is repeated one thousand times to obtain measurement statistics and the resulting averaged time-traces are presented in Fig. 2c. An exponential decay of the presence of the electron in the dot as a function of the time spent at working point B is observed in the experimental data corresponding to a tunnelling time close to 1 second as indicated by the green line. This gate pulsing sequence is then repeated by adding a burst of microwave (MW) to the IDT with a pulse length of several tenths of nanoseconds and which is applied one hundred milliseconds after the system is brought into position B. The MW burst creates a moving quantum dot, which lifts the electron, initially trapped in the left quantum dot, above the tunnel barrier and drags it out of the quantum dot. This results in a jump in the QPC current as shown by the red line.

In order to demonstrate that the electron has been loaded into a moving quantum dot and not expelled into the reservoir, it is essential to detect the coincidence between events when the electron leaves the single electron source (left dot) and when it is trapped in the single electron detector (right dot). This is realized by a second voltage pulse sequence on the right dot: when the single electron source is brought in position B, the detector dot is armed by pulsing its gates to working point B' where the steady state is the zero electron state and the coupling to the channel is large. At this working point both QPC traces are recorded simultaneously. No charge variation is observed during the first 50 ms where the system is kept in position B. A MW pulse is sent on a time-lag of 50 ms. After the recording, the detector is reinitialized to zero electron at working point A' where the captured electron can tunnel efficiently into the reservoir. Typical single shot read out curves are presented in Fig. 3a-d. Coincidences are observed between events when an electron leaves the source quantum dot and an electron is detected in the receiver quantum dot within the same timeslot (Fig. 3a). These events correspond to the scenario where one electron has been loaded in the electron source (left dot), is then transferred in the quantum channel (the moving quantum dots) and is received in the detector (right dot). It is worth noting that in comparison to photon detectors here the electron still exists after detection. A set of experiments described in Fig. 3 allows to fully characterize the high quantum efficiency of both the single electron source and the single electron detector observed in the experiment: 96 % for the single electron source and 92 % for the single electron detector (see Fig. 3e).

In quantum dots, it is possible to load not only one but two electrons. By waiting long enough[17], the two electrons will be in a singlet state at zero magnetic field and are hence entangled in the spin degree of freedom. The ability to separate the two electrons and to bring only one of them to the second quantum dot is of potential interest in order to transfer quantum information and is at the essence of the quantum teleportation protocol[2, 18, 19, 20]. In analogy with photons, this is the equivalence of a two-photon entangled source[21]. Moreover, in contrast to a photon detector, the electron detector can discriminate easily whether one, two or more electrons left the single electron source and are captured in the single electron detector (see Fig. 2a). The protocol consists in loading the left dot with exactly two electrons by moving the gates Vb and Vc into the two-electron regime of the stability diagram. The quantum dot is then tuned towards the working point where loading of the moving quantum dots is possible (point B). Different possibilities for the emission of electrons into the quantum channel are observed. Indeed, when starting with exactly two electrons in the source dot, one can achieve that either exactly one or both electrons are emitted from the source and received in the detector dot, as shown by the single shot traces for QPC detection of the two dots (see Fig. 4a-d). The probability of each event varies with the working voltage point B. For very negative gate voltage Vc, about half of the time the two electrons are separated, meaning that only one electron is transferred, and the other half of the time the two electrons are transported (see Fig. 4e). For the events where the two electrons leave the dot, the electrons are most probably loaded into two different moving quantum dots. More interestingly, when pulsing the gate Vc more positively, a regime can be realized where only one of the two electrons of the left dot is efficiently emitted and consequently captured by the right dot (see Fig. 4e). In this case, the probability of sending the two electrons is dramatically reduced to below 3 % and the probability to effectively separate the two electrons approaches 90 %.

In order to use single electron transfer in quantum operations using spin qubits, one has to demonstrate that the coherence of the electron spin after the electron transfer is preserved. Measurement and coherent manipulations of electron spins can be straightforwardly implemented in our set-up and the spin coherence time $T_2$* of an ensemble of electrons stored in SAW assisted moving quantum dots has been shown to be as long as 25 ns[13]. A necessary condition to investigate coherent transport of a single electron spin is to be able to trigger the electron transfer within a timescale, which is short compared with $T_2$*. Indeed, a microwave pulse of a length of 250 ns corresponds to about 700 moving quantum dots and the above described experiments demonstrate the ability to load the electron into one of the moving quantum dots produced by each SAW microwave burst. In this last section, we demonstrate that the number of minima of the microwave burst in which the electron is loaded can be narrowed down to two. For this purpose, the single electron source voltage sequence is slightly modified. After the charging of the quantum dot, the system is brought to position B (see Fig. 2a) slightly on the more negative side with respect to Vc and the duration of the microwave pulse is shortened to a minimum value of 65 ns. At this voltage position, the barrier height to the quantum channel is increased and the transfer probability of an electron into the quantum channel is as low as 5% when excited with the SAW microwave burst. To trigger the single electron transfer, a 1ns-pulse on gate Vc with a positive value (voltage position C in Fig. 2a) is added to this sequence. In Fig. 4f, the evolution of the number of events where one electron leaves the single electron source and one electron is detected in the single electron detector ($N_{1001}$) is plotted as a function of the delay between the 1 ns-gate pulse and the 65 ns microwave burst. High transfer probabilities reaching 90 % are only observed for time delays around 765 ns, corresponding to the propagation time of the surface acoustic wave from the IDT to the dot region. Taking into account the pulse length of the gate and the distance between two minima of the SAW, only two moving quantum dots can then be the hosts of the transported electron during the gate pulse as schematically indicated in Fig. 4g. This demonstrates the ability to load on demand and in a very reproducible manner one of the two minima of the train of moving quantum dots with a single electron during the 1ns-gate pulse. Using a faster arbitrary waveform generator should allow loading the electron on demand into the same moving quantum dot.

The above experiments represent the first milestone towards a new experimental platform to realize quantum optics with flying electrons implemented in gated 2DEG heterostructures and transported by surface acoustic waves. High quantum efficiency of both the single electron detector and the single electron source are demonstrated and potentially enable to measure all moments of the electron correlations[22]. In comparison with other implementations in similar systems, the propagating electron is physically isolated from the other conduction electrons of the heterostructure. In bringing together two propagating quantum buses separated by a tunnel barrier, a beam splitter for flying electrons can be implemented[23,24] and Hanbury Brown and Twiss type experiments where stronger Coulomb interactions between electrons could be realized. Future experiments should allow coherent spin transfer and give new insight on the feasibility of quantum teleportation protocols and on the potential scalability of spin qubits.

## METHODS

The device is defined by Schottky gates in an n-AlGaAs/GaAs 2DEG based heterostructure (2DEG: $\mu \cong 10^6 \, \mathrm{cm^2/Vs}$, $n_s \cong 1.4 \times 10^{11} \mathrm{cm^{-2}}$, depth = 90 nm) using standard split-gate techniques. The charge configuration of both dots are measured via the conductance of both QPC by biasing it with a DC voltage of 300μV and the current is measured with an IV-converter with a bandwidth of 1.4 kHz. The voltage on each gate can be varied on a timescale down to μs. In addition, the gate Vc controlling the coupling between the left dot and the 1D channel is connected to a homemade bias-tee to allow nanosecond manipulation of the dot potential by means of an arbitrary function generator (Tektronix AWG 5014). The IDT, which is placed at a distance of approximately 2 mm to the left of the sample, is made of 70 pairs of lines, 70 μm in length and 250 nm in width with a repetition of 1 μm. The IDT is oriented perpendicular to the direction of the 1D-channel defined along the crystal axis [110] of the GaAs wafer and has a frequency bandwidth of approximately 20 MHz. By applying a radio-frequency burst to the IDT, a surface acoustic wave is generated which travels with a speed of approximately 3000 m/s across the GaAs crystal. Due to the piezo-electric properties of GaAs, the generated moving electrostatic potential can drag electrons along the propagation direction of the SAW[13] (see supplementary material).

**Acknowledgements**

We acknowledge technical support from the "Poles Electroniques" of the "Department Nano et MCBT" from the Institut Néel as well as Yannick Launay and Pierre Perrier for technical support. M.Y. acknowledges financial support by Grant-in-Aid for Young Scientists A (no. 20684011) and ERATO-JST (080300000477). S.T. acknowledges financial support from Special Coordination Funds for Promoting Science and Technology (NanoQuine), JST Strategic International Cooperative Program, MEXT KAKENHI "Quantum Cybernetics" and IARPA project "Multi-Qubit Coherent Operations" through Harvard University.A.D.W. acknowledges expert help from PD Dr. Dirk Reuter and support of the DFG

SPP1285 and the BMBF QuaHLRep 01BQ1035. C.B. acknowledges financial support from CNRS (DREI) - JSPS (no. PRC 424 and L08519). T.M. acknowledges financial support from Marie-Curie ERG 224786.

We are grateful to the Nanoscience Foundation of Grenoble, for partial financial support of this work.

**Figure 1 Experimental device and measurement set-up** SEM image of the single electron transfer device and schematic of the experimental setup. Two quantum dots, which can be brought into the single electron regime, are separated by a 3µm long 1D-channel as shown in Fig. 1. Each quantum dot is capacitively coupled to a close by quantum point contact (QPC) that is used as an electrometer [9]. By applying a 65 ns long microwave burst on the interdigitated transducer (IDT, see methods for details), a train of about 150 moving quantum dots are created in the 1D channel. Vc is connected to a homemade bias-tee to allow nanosecond manipulation of the dot potential.

**Figure 2 Stability diagrams of the two quantum dots and charge detection**

**a,b,** Stability diagram of the left (a) and the right dot (b) obtained via charge detection by varying the gate Vb(b') (gate controlling the coupling to the reservoir) and Vc(c') (gate controlling the coupling to the channel). Sweeps in Vb(b') are fast and are carried out within 1s from +0.15 V to -0.15 V (3ms per point). When the barrier height is made higher (Vb more negative), metastable charge states with timescales longer compared to the Vb(b') sweep time are observed. In the very negative Vb part of the diagram for the right dot, the electrons will finally tunnel out. When the sweep direction of the Vb is reversed these charge detection steps are absent. **Inset a,** Schematic description of the dots+channel electrostatic potential applied by the gates to the electron at different points in the stability diagram (see text)**. c,** Average QPC time trace along the voltage sequence of the single electron source. Without the MW burst applied on the IDT, we observe a lifetime for the metastable 1 electron charge state of 700ms. Applying a MW burst, the electron in the metastable state is forced to quit the quantum dot with very high probability.

**Figure 3 Coincidence between emission and detection of a single electron**

Coincidence between the two single shot QPC time-traces at voltage working point B and B' corresponding to the different events: **a,** N1001 **b** N1000 **c,** N1100 **d**, N0000. The position in time of the RF burst is indicated by the black arrow. At this specific time, the small peak (dip) observed on time traces are the result of the SAW induced enhancement (reduction) of the QPC current. Notation $N\alpha\beta\gamma\delta$ corresponds to the number of events with $\alpha$ ($\beta$) electrons in the source dot before (after) the MW burst and to $\gamma$ ($\delta$) electrons in the receiver dot before (after) the MW burst. When one index is replaced by x, the corresponding output result is disregarded. Event N1000 corresponds to the situation where the electron has been transferred from the source to the detector and is immediately kicked out of the detector dot by the same RF burst and hence not detected. Events where $\beta+\delta > \alpha+\gamma$, are called a "bad" event. **e,** Summary table of the different events over 10001 traces for different experimental situations: (1) an electron is loaded in the source and a RF burst is applied, (2) an electron is loaded in the source dot and no RF burst is applied, (3) no electron is loaded in the source dot and a RF burst is applied, (4) an electron is loaded in the source dot close to the degeneracy point (DP) with a probability of tunnelling into the dot equal to 1/2 and a RF burst is applied, (5) an electron is loaded in the source dot close to the DP with a probability of tunnelling into the dot equal to 1/6 and a RF burst is applied. The presented set of data allows to conclude that the transfer of a single electron is achieved and to determine the quantum efficiency of the source (detector) equal to 96% (92%).

**Figure 4 Coincidence between emission and detection of two electrons a,** Coincidence between the two single shot QPC time-traces at voltage working point B and B' corresponding to the different events: **a,** N2100 **b** N2101 **c,** N2001 **d**, N2002. **e,** Summary table of the different events over 1005 traces for different experimental situations: (1) two electrons are loaded in the source dot and a RF burst is applied with Vc=-0.388 V. (2) two electrons are loaded in the source dot and a RF burst is applied with Vc=-0.322 V. The presented set of data allows to conclude that the two electrons from a singlet state in a single dot can be separated into the two distant dots with an efficiency close to 90%. **f, Triggered nanosecond electron transfer. f,** Evolution of the number of the N1001 and N10xx events as a function of the delay between the 1ns-gate pulse and the 65ns-microwave burst when a single electron is loaded into the single electron source. **g**, Schematic description of the timing sequence between 1ns-gate pulse and MW busrt applied on the IDT.

**Fig1**

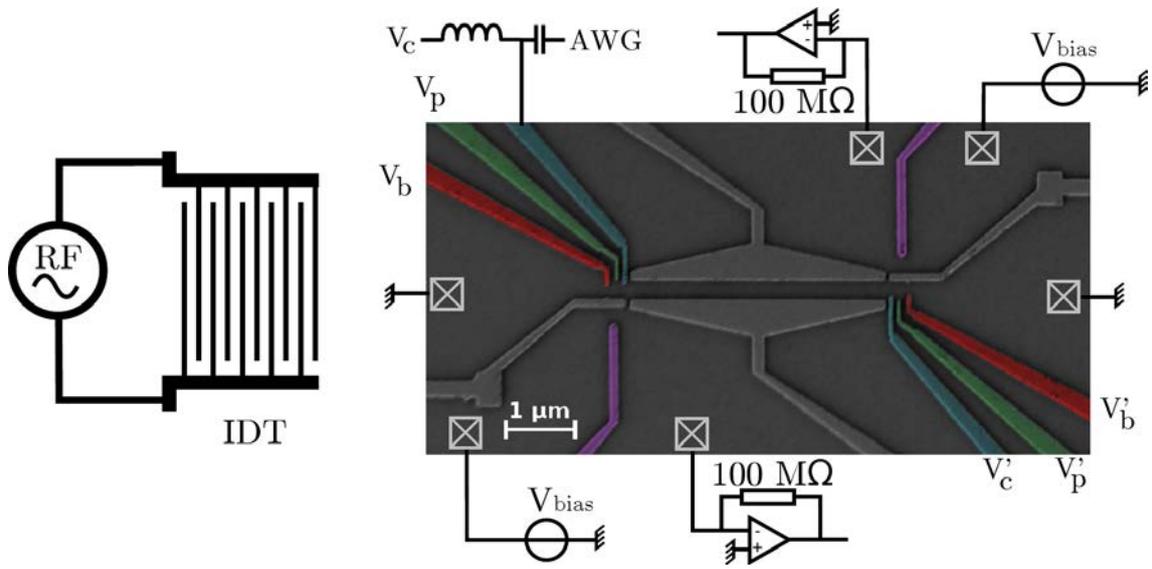

**Fig2**

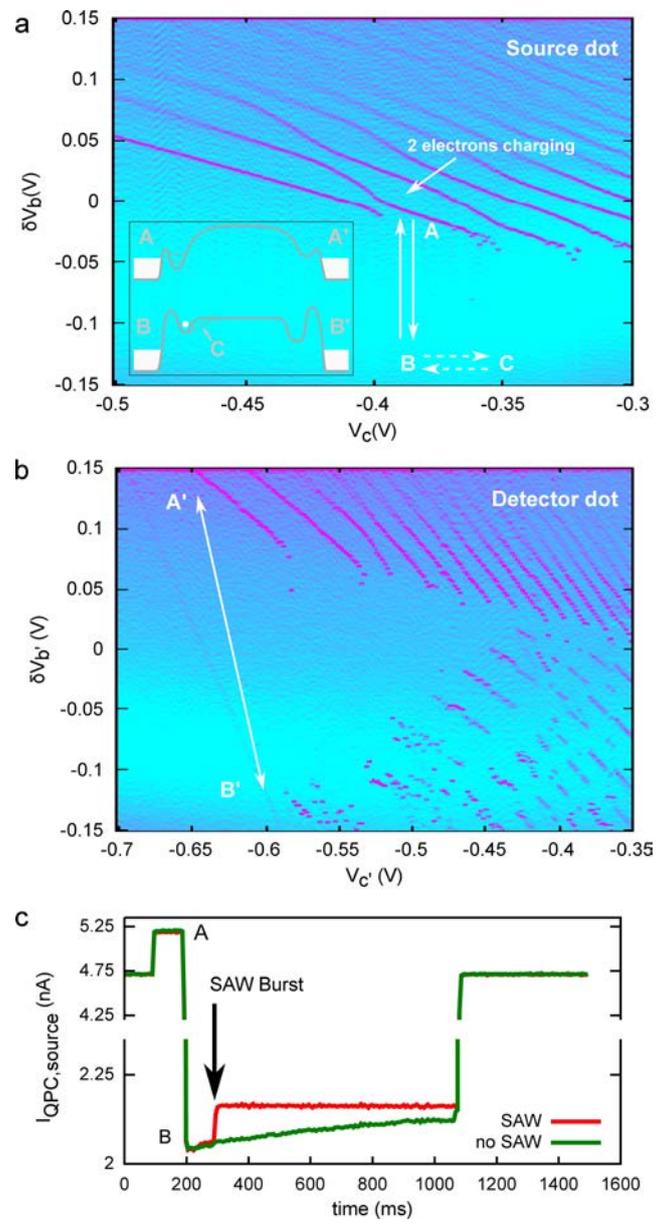

**Fig3**

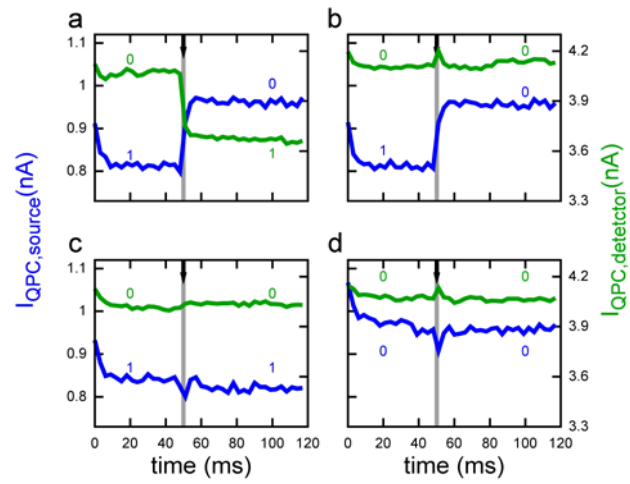

|   | (1) | (2) | (3) | (4) | (5) |
|---|---|---|---|---|---|
| $N_{1xxx}$ | 9841 | 10001 | 16 | 5154 | 1462 |
| $N_{10xx}$ | 9408 95.6% | 0 0% | 15 94% | 4954 96.1% | 1395 95.4% |
| $N_{100x}$ | 9128 | 0 | 14 | 4807 | 1349 |
| $N_{1001}$ | 8393 91.9% | 0 0% | 14 100% | 4417 91.9% | 1244 92.2% |
| $\sum N\alpha\beta\gamma\delta$ $_{\beta-\alpha < \delta-\gamma}$ | 0 | 1 | 0 | 0 | 0 |

**Fig4**

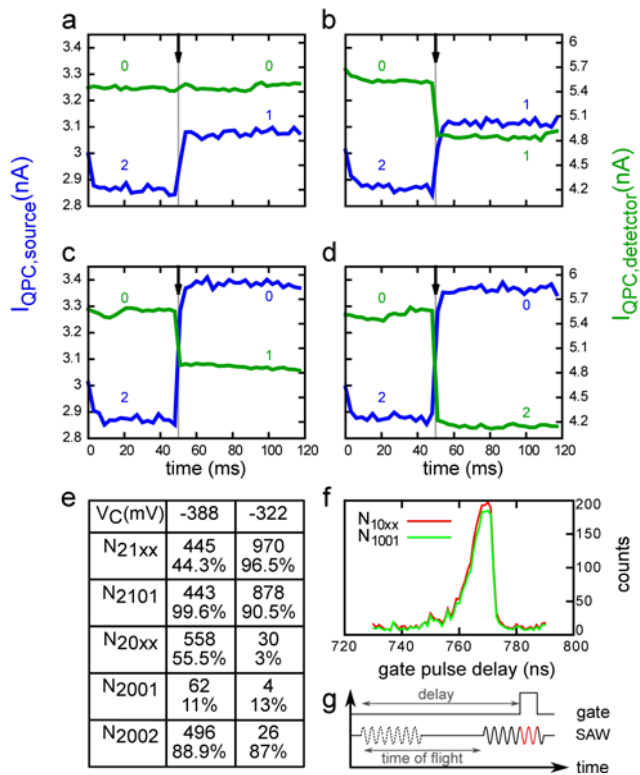

# Supplementary online material to

# Electrons surfing on a sound wave as a platform for quantum optics with flying electrons

Sylvain Hermelin[1], Shintaro Takada[2], Michihisa Yamamoto[2,3], Seigo Tarucha[2,4],

Andreas D. Wieck[5], Laurent Saminadayar[1,6], Christopher Bäuerle[1], and Tristan

Meunier[1]

[1]*Institut Néel, CNRS and Université Joseph Fourier, 38042 Grenoble,*

*France* [2]*Department of Applied Physics, The University of Tokyo, Tokyo, 113-8656,*

*Japan,*

[3]*ERATO-JST, Kawaguchi-shi, Saitama 331-0012, Japan,*

[4]*ICORP (International Cooperative Research Project) Quantum Spin Information*

*Project, Atsugi-shi, Kanagawa, 243-0198, Japan,*

[5]*Lehrstuhl für Angewandte Festkörperphysik, Ruhr-Universität Bochum,*

*Universitätsstraße 150, 44780 Bochum, Germany*

[6]*Institut Universitaire de France, 103 boulevard Saint-Michel, 75005 Paris, France*

**Calibration of the SAW excitation**

The SAW transducer is placed approximately 2mm away from the center of the sample

and were realized by electron beam lithography and standard lift-off technique using a

titanium-gold metallic layer. The distance between adjacent fingers is 500 nm (half

wave length). RF characterization at room temperature showed a resonance frequency

around 2.6 GHz. Characterization at 100 mK was realized in two steps. First the

conductance of the left quantum dot was measured using a low frequency lock-in

technique in voltage bias regime. Fig. S1 shows the measured current when the RF

signal is applied to the IDT with a duty cycle 1:50 at 1 dBm, while scanning the RF frequency and the plunger gate of the dot. A clear resonance can be seen at 2.6326 GHz.

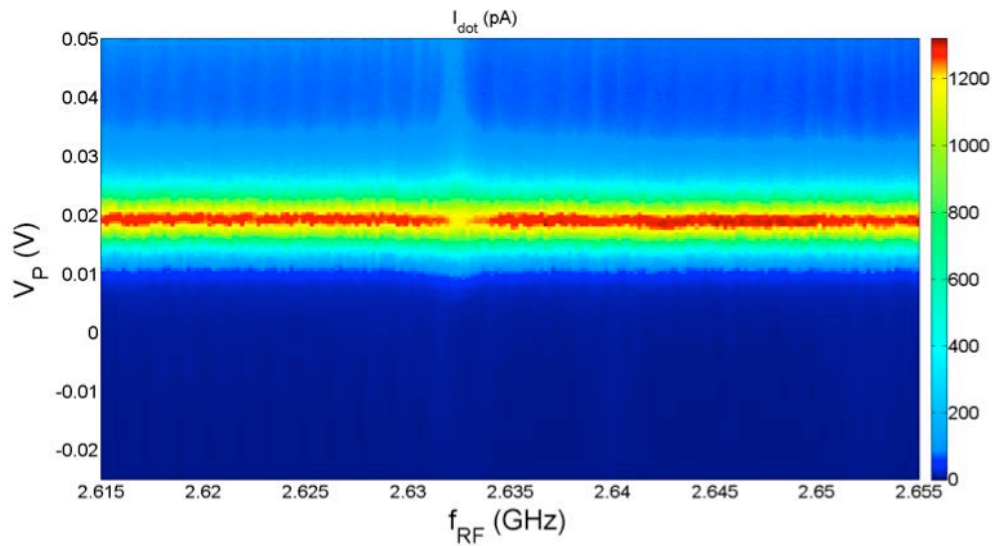

**Fig. S1 IDT characterization** Lock-in current through a quantum dot under SAW irradiation (1ms:50ms duty cycle) while scanning through a Coulomb peak. The SAW only acts when the IDT is at its resonance frequency of 2.6326 GHz.

The working frequency was then fixed at this value. Fig. S2 shows the influence of the RF power versus the plunger gate voltage at duty cycle 1:1, which allows to calibrate the SAW amplitude against the RF power at the source and yields an amplitude A[eV] = 2/25*10^(P[dBm]-23)/20.

For the single and two-electron transfer, the RF power applied on the IDT was 14 dBm.

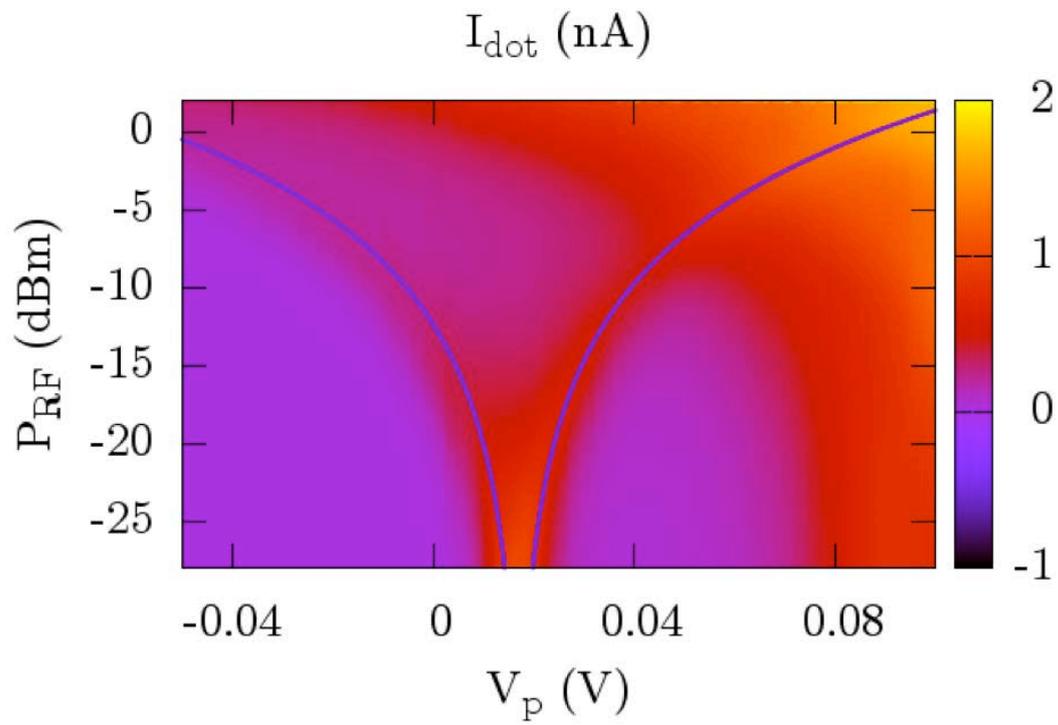

**Fig. S2 IDT calibration** Current amplitude through a quantum dot under continuous SAW irradiation. The SAW amplitude splits the Coulomb peak by increasing the amplitude of the SAW electrical potential[1].

**Quantized SAW current**

In our set-up, the region between the two electrodes, which define the 1D channel is fully depleted. As a consequence, direct linear electron transport from one end of the channel to the other is blocked since the Fermi energy lies below the potential induced by the gates. By applying a microwave excitation to the IDT, a SAW induced acoustic current can be generated. This has been demonstrated in pioneering experiments by the Cavendish group[2]. Indeed, when SAWs are excited at the surface of the heterostructure, a moving electrostatic potential landscape is generated within the 2DEG. When confined to a one-dimensional channel, moving quantum dots can hence be realized. Electrons are trapped in a minimum of the SAW potential and are dragged along the minimum of the electric field potential generated by the moving quantum dots. Therefore, electron transport from one end of the channel to the other is restored and quantized conductance corresponding exactly to an integer number of electrons per moving quantum dots has been observed[2].

The quantized SAW current corresponding to the transfer of an integer number of electrons per potential minimum in the moving dots has also been observed in our SAW set-up. This current quantization, however, requires a potential profile, in which exactly an integer number of electrons are captured by each moving potential minimum at the entrance of the 1D channel. In our device used for the demonstration of the single electron transfer between the two quantum dots, the 1D channel was made very narrow (0.3 μm) and the SAW potential could not catch exactly a single electron at the entrance of the 1D channel. Quantized conductance was instead observed by employing a wider one-dimensional channel. Fig. S3 shows such quantized current through a 0.7 μm wide channel observed at Helium temperature (~4K).

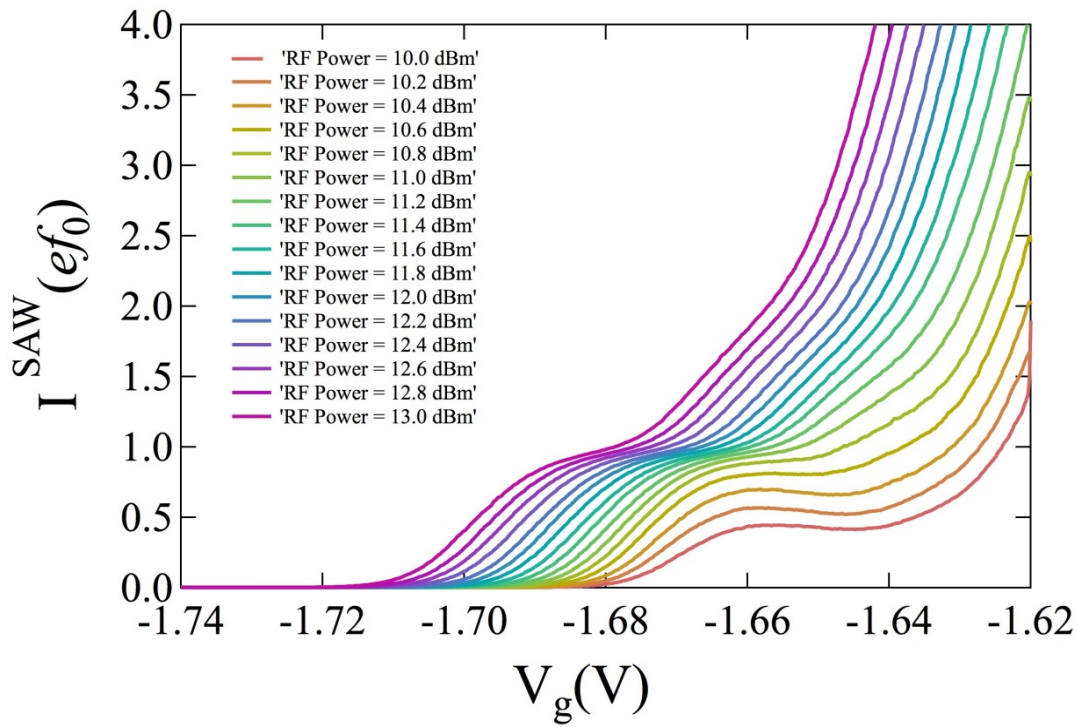

**Fig. S3 Quantized SAW current** Current through a 0.7 μm wide and 0.8 μm long quantum wire under continuous SAW irradiation as a function of the gate voltage Vg of the quantum wire for different RF power at 4 K. Current is plotted in units of $ef_0$, with $f_0$ = 2.696 GHz.